\documentclass[aps,prb,reprint,superscriptaddress,showkeys,showpacs,amssymb,amsmath]{revtex4-1}

\usepackage{graphicx}% Include figure files
\usepackage[latin1]{inputenc} % support Umlaute
\usepackage[USenglish]{babel} %hyphenation in american english
\usepackage{hyperref}
\hypersetup{
    colorlinks=true,        % false: boxed links; true: colored links
    linkcolor=red,          % color of internal links
    citecolor=blue,        % color of links to bibliography
    filecolor=magenta,      % color of file links
    urlcolor=blue           % color of external links
}
\usepackage{siunitx}

\begin{document}

\title{Interplay between the electrical transport properties of GeMn thin films and Ge substrates}

\author{N. Sircar}
  \affiliation{Walter Schottky Institut, Technische Universit\"at M\"unchen, Am Coulombwall 4, D-85748 Garching, Germany}
\author{S. Ahlers}
  \affiliation{Walter Schottky Institut, Technische Universit\"at M\"unchen, Am Coulombwall 4, D-85748 Garching, Germany}
\author{C. Majer}
	\affiliation{Walter Schottky Institut, Technische Universit\"at M\"unchen, Am Coulombwall 4, D-85748 Garching, Germany}
\author{G. Abstreiter}
	\affiliation{Walter Schottky Institut, Technische Universit\"at M\"unchen, Am Coulombwall 4, D-85748 Garching, Germany}
\author{D. Bougeard}
  \affiliation{Walter Schottky Institut, Technische Universit\"at M\"unchen, Am Coulombwall 4, D-85748 Garching, Germany}
  \affiliation{Institut f\"ur Experimentelle und Angewandte Physik, Universit\"at Regensburg, D-93040 Regensburg, Germany}

%
%%%%%%%%%%%%%%%%%%%%%%%%%%%%%%%%%%%%%%%%%%%%%%%%%%%%%%%%%%%%%%%%%%%%%%%%%%%%%%%%%%%%%%%%%%%%%%%
%%%%% abstract
%%%%%%%%%%%%%%%%%%%%%%%%%%%%%%%%%%%%%%%%%%%%%%%%%%%%%%%%%%%%%%%%%%%%%%%%%%%%%%%%%%%%%%%%%%%%%%%
\begin{abstract}
We present evidence that electrical transport studies of epitaxial p-type GeMn thin films fabricated on high resistivity Ge substrates are severely influenced by parallel conduction through the substrate, related to the large intrinsic conductivity of Ge due to its small bandgap. Anomalous Hall measurements and large magneto resistance effects are completely understood by taking a dominating substrate contribution as well as the measurement geometry into account. It is shown that substrate conduction persists also for well conducting, degenerate, p-type thin films, giving rise to an effective two-layer conduction scheme. Using n-type Ge substrates, parallel conduction through the substrate can be reduced for the p-type epi-layers, as a consequence of the emerging pn-interface junction. GeMn thin films fabricated on these substrates exhibit a negligible magneto resistance effect. Our study underlines the importance of a thorough characterization and understanding of possible substrate contributions for electrical transport studies of GeMn thin films.
\end{abstract}

\maketitle
\section{Introduction}
In the past years the emerging field of spintronics has led to the search for novel materials exhibiting ferromagnetic and semiconducting properties at the same time, since such ferromagnetic semiconductors would allow the integration of new application schemes into established semiconductor technologies. GeMn seems to be a very promising candidate in the class of ferromagnetic semiconductors for its compatibility with the mainstream Silicon technology. Recent works consistently demonstrated the possibility of preparing GeMn thin films by molecular beam epitaxy (MBE) without secondary phase separation, but with the strong tendency to the formation of Mn rich nanometer sized clusters.\cite{Bou2006, Jam2006, Bougeard2009, Li2007} These clusters exhibit a Curie temperature about or even above room-temperature (RT) which would be important for device applications. In an early work it was claimed that magnetic properties may be controlled through electric gating, suggesting that charge carriers mediate the magnetic exchange interactions.\cite{Par2002} Therefore the electrical properties in an external magnetic field, i.e. the magneto resistance (MR) and the anomalous Hall effect (AHE) are considered an important fingerprint of a magnetic semiconductor. However, in the same degree as there is a consistent picture of the nanostructure of GeMn thin films, their fingerprint in magneto transport measurements still lacks such a coherent description in literature. For example, MR effects reaching from several thousand percent being positive\cite{Jam2006} to a few percent being negative\cite{Li2007, Li2005a, Zhou2010} have been reported. Similarly, Hall effect measurements sometimes yield a large contribution of the AHE on the one hand,\cite{Li2005a, Deng2009} but also a diminishing contribution washed out by the ordinary Hall effect on the other.\cite{Zhou2009} In many cases the interpretation of these results interestingly does not correlate with the sample magnetization, particularly regarding its saturation and hysteresis effects, which are often absent in transport measurements. Recently it was pointed out by \citet{Zhou2009} that part of these reports may be understood in a regime of parallel conduction of two charge carrier types, owing to the role of Mn as a deep two-level acceptor in Ge, without being related to the magnetic nature of GeMn thin films in itself.

In this study, we would like to highlight the possibility that the peculiar transport properties observed in some GeMn transport studies might not be caused by the alloying of Ge with Mn at all, but by parallel conduction through the employed substrates. While high purity substrates of semiconductors like Si or GaAs exhibit RT resistivities greater than \SI{e3}{\ohm\centi\metre}, this is not the case for Ge substrates. Since intrinsic conduction in Ge already becomes important around RT due to its small bandgap, RT resistivities are intrinsically limited around \SI{50}{\ohm\centi\metre}.\cite{Mor1954a} This upper limit is already reached with impurity concentrations as low as \SI{e13}{\per\cubic\centi\metre}.\cite{Pri1953} Substrates with higher purity are commercially not available.  We will show that the electrical properties of epitaxial GeMn thin films fabricated by solid-source MBE on such high purity Ge substrates can severely be influenced by parallel conduction through the substrate. To furthermore demonstrate the effects of parallel conduction, we studied a system of non-magnetic, degenerately doped Ge:B epitaxial layers grown on these high purity Ge substrates. Some of these results have a remarkable resemblance to previously published data on magnetic GeMn thin films,\cite{Riss2009,Zhou2009} although our Ge:B films do not show any sign of magnetism other than common diamagnetism. We will give a two-layer model accounting for the parallel conduction through the substrate, which sufficiently well describes the experimental magneto transport results in those types of thin films. 

\section{Experimental}
\begin{figure}
	\centering
		\includegraphics{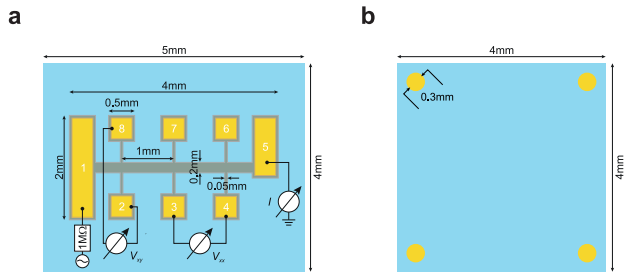}
	   \caption{\textbf{(a)} Scaled schematic of samples with Hall bar mesa. The principle setup for measuring the longitudinal ($V_{xx}$) and Hall ($V_{xy}$) voltage as well as total current ($I$) is also indicated. \textbf{(b)} Geometry of samples where the van der Pauw method has been employed.\label{fig:HB_vdP}}
\end{figure}
The investigated samples were fabricated with solid-source MBE under ultra high vacuum conditions at a base pressure of \SI{5e-11}{\milli\bbar}. We used high resistivity Ge(001) substrates with a RT resistivity larger than \SI{40}{\ohm\centi\metre} and thickness of approximately \SI{500}{\micro\meter}. These substrates are specified to exhibit n-type conduction, due to Antimony impurities dissolved into the Ge crystal during its fabrication. However, we would like to note that the vendors specification of this n-type conduction is only true about RT. In fact, the substrate undergoes a transition to p-type conduction below RT. This indicates the presence of a majority of residual acceptor-like impurities. Hence, the substrate may suffer from considerable auto compensation. At RT, the conduction behavior is dominated by intrinsic charge carriers, primarily by electrons due to their smaller effective mass, and is therefore n-type. 

Prior to growth of all thin films a \SI{80}{\nano\metre} thick, undoped Ge buffer layer was deposited. The GeMn sample was grown by co-deposition of Mn and Ge at a Ge growth rate of $r_\textrm{Ge} = \SI{0.08}{\angstrom\per\second}$ and at constant substrate temperature $T_\textrm S = \SI{60}{\celsius}$ to avoid the formation of intermetallic secondary phases. The film thickness amounts to \SI{200}{\nano\meter} with a total Mn concentration of \SI{5}{\percent}. A thorough characterization of the structural and magnetic properties of this sample may be found elsewhere.\cite{Bou2006}

For comparison, we also fabricated a non-magnetic p-type Ge thin film using a Boron effusion cell. This sample was fabricated at $r_\textrm{Ge} = \SI{0.3}{\angstrom\per\second}$ and $T_\textrm S = \SI{360}{\celsius}$ with a thickness of \SI{200}{\nano\metre}. The B concentration of \SI{5e19}{\per\cubic\centi\metre} was chosen to be well above the insulator-to-metal transition.\cite{Fritzsche1959}

For transport measurements an approximately \SI{450}{\nano\meter} deep Hall bar mesa was defined by standard lithography methods and wet-chemical etching. Geometrical details are iven in Figure~\hyperref[fig:HB_vdP]{\ref{fig:HB_vdP}(a)}. The longitudinal and Hall resistance, $R$ and $R_{xy}$, were determined by applying a current $I$ along the Hall bar and measuring the longitudinal and transversal voltages, $V_{xx}$ and $V_{xy}$, in a standard, quasi-DC lock-in setup using an additional \SI{1}{\tera\ohm} input impedance voltage amplifier before the lock-in. As will be introduced in section~\ref{sec:vdPgeometry}, some samples were also investigated via the van der Pauw method in the geometry show in Fig.~\hyperref[fig:HB_vdP]{\ref{fig:HB_vdP}(b)}.

Temperature dependent resistance measurements without applied magnetic field were performed with a heatable sample stick inserted in a liquid Helium dewar. Field dependent measurements were performed in a variable temperature magnet-cryostat, with the magnetic field applied perpendicular to the sample surface.

\section{Results and Discussion}
\subsection{Non-degenerate GeMn on high resistivity Ge substrates}
\begin{figure}
	\centering
		\includegraphics{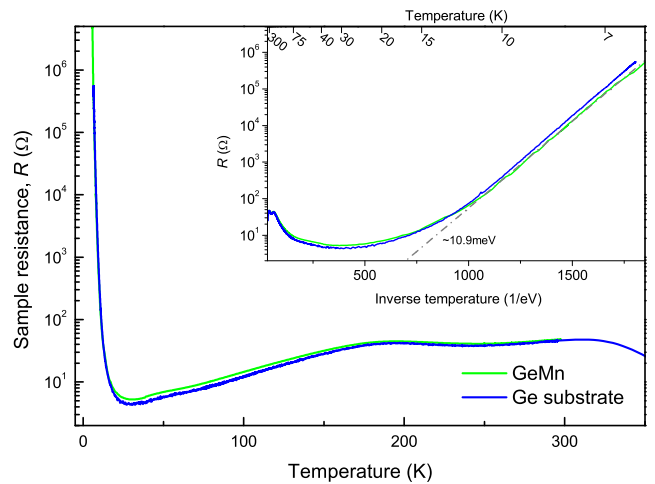}
	   \caption{Sample resistance versus temperature for the GeMn sample (green) and the Ge substrate reference (blue). Inset: Same data as function of inverse temperature. A straight line (grey) corresponding to an activation energy of \SI{10.9}{\milli\electronvolt}
	    can be fitted to the extrinsic freeze-out.\label{fig:rho(T)_GeMn_hrGeSub}}
\end{figure}

Figure~\ref{fig:rho(T)_GeMn_hrGeSub} depicts the sample resistance of the bare high resistivity Ge substrate sample as a function of temperature. We can identify the three distinct regions well known for non-degenerate semiconductors, i.e. the freeze-out of extrinsic charge carriers, the extrinsic and the onset of the intrinsic range.

Also shown in Fig.~\ref{fig:rho(T)_GeMn_hrGeSub} is the resistance measurement of the GeMn thin film grown on the high resistivity Ge substrate. When comparing the two samples, we notice that the resistance of both samples is of the same order of magnitude and has a very similar temperature dependence. This becomes more evident in an Arrhenius plot of the resistance depicted in the inset of Fig.~\ref{fig:rho(T)_GeMn_hrGeSub}. Both curves exhibit the same linear slope in the extrinsic freeze-out regime, which corresponds to a thermal activation energy of $E_\mathrm{A}=\SI{10.9}{\milli\electronvolt}$ for the dopant impurities.\cite{[{The temperature dependence of the mobility has been neglected. Due to auto compensation we fitted the data with an $exp{\frac{E_A}{k_{B}T}}$ law, see for example }]Bla1962} This is in good agreement with the activation energy of shallow impurities in Ge. It does not correspond with the activation energy of Mn in the Ge, which is expected to be a two-level deep band gap acceptor with $E_\mathrm{A} = \SI{160}{\milli\electronvolt}$ and \SI{370}{\milli\electronvolt}, respectively.\cite{Woo1955} It rather seems that in both samples the residual impurities dissolved in the Ge substrate dominate the measurements.

\begin{figure}
	\includegraphics{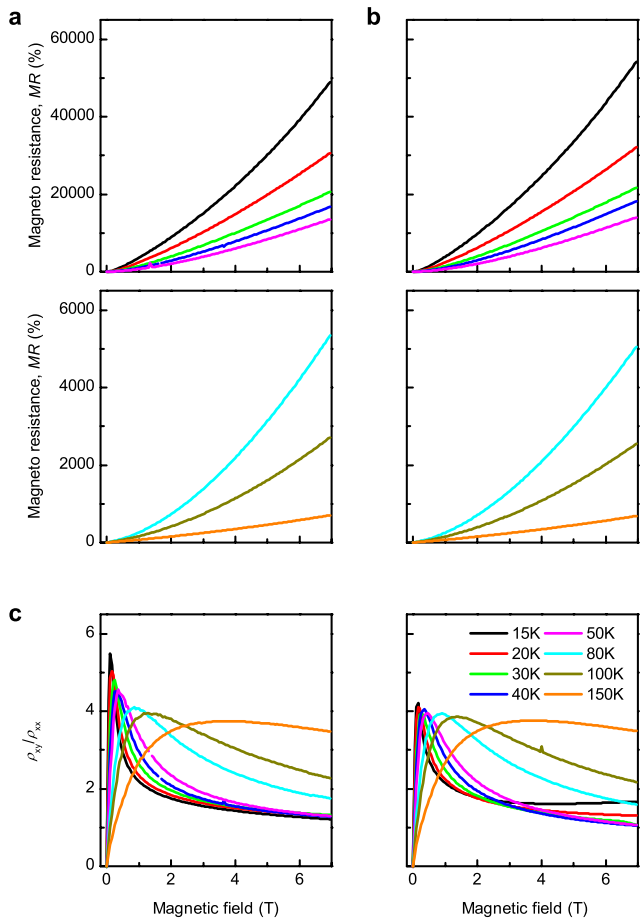}
	 \caption{\textbf{(a)} Transversal MR for various temperatures for the GeMn sample. \textbf{(b)} MR for the bare high resistivity Ge substrate. \textbf{(c)} Tangent of the Hall angle for (left) the GeMn sample and (right) the Ge substrate. The color code is the same for all panels.	\label{fig:MR_Hallangle_Sub_GeMn_HB}}
\end{figure}

Figure~\hyperref[fig:MR_Hallangle_Sub_GeMn_HB]{\ref{fig:MR_Hallangle_Sub_GeMn_HB}(a)} shows the MR effect of the GeMn sample versus external magnetic field $B$ for various temperatures, calculated according to the convention
\begin{equation}
	\mathrm{MR}\left[\%\right] = \frac{R(B)- R(0)}{R(0)}\cdot 100.
\end{equation}
The MR effect is positive and exhibits a parabolic-like dependence for weak fields tending towards a linear dependence at higher fields, without any signs of saturation. For higher temperatures the MR effect gradually decreases in its magnitude. Similar results on the MR effect in GeMn have already been reported.\cite{Jam2006} 

The orbital MR of the semi-classical Boltzmann transport theory can not be responsible for the positive MR effect depicted in Fig.~\hyperref[fig:MR_Hallangle_Sub_GeMn_HB]{\ref{fig:MR_Hallangle_Sub_GeMn_HB}(a)}, since the order of magnitude of the MR is too large. In fact, an orbital MR effect would also not explain the non-saturating character of the observed MR at large fields.\cite{Hurd1972} A connection of the MR to the magnetic nature of the GeMn epi-layer can be ruled out for the same reason, as its magnetization saturates at fields about \SI{2}{\tesla}.\cite{Bou2006} \citet{Jam2006} proposed the occurrence of a geometrically enhanced MR effect\cite{Solin2000} to account for the large magnitude and the linear increase at high fields of the MR in their GeMn sample, stemming from the presence of highly conducting Mn-rich inclusions. Such an inhomogeneous semiconductor can indeed exhibit extremely large, non-saturating MR.\cite{Parish2005,Parish2003,Hu2007} However, we obtain essentially equal results for the magnitude as well as field and temperature dependence of the MR of the bare Ge substrate, as shown in Fig.~\hyperref[fig:MR_Hallangle_Sub_GeMn_HB]{\ref{fig:MR_Hallangle_Sub_GeMn_HB}(b)}. 

The left panel in Figure~\hyperref[fig:MR_Hallangle_Sub_GeMn_HB]{\ref{fig:MR_Hallangle_Sub_GeMn_HB}(c)} depicts the tangent of the Hall angle of the GeMn sample, defined as $\rho_{xy}$/$\rho_{xx}$ with $\rho_{xy}$ and $\rho_{xx}$ being the Hall and longitudinal resistivities, respectively. The Hall angle gives a more direct estimate of possible magnetization induced contributions to the ordinary Hall effect than the common Hall curve. The Hall angle increases steeply with field, tending towards a saturation at higher fields. Similar results were found by other groups for the GeMn material system and were either attributed to a magnetization induced AHE\cite{Jam2006} or related to the multiple Mn acceptor energy states leading to an effective two-band like conduction.\cite{Zhou2009} Our undoped, non-magnetic Ge substrate exhibits the same Hall angle behavior as can be seen in the right panel of Fig.~\hyperref[fig:MR_Hallangle_Sub_GeMn_HB]{\ref{fig:MR_Hallangle_Sub_GeMn_HB}(c)}.

The data presented on the GeMn thin film in Fig.~\ref{fig:rho(T)_GeMn_hrGeSub} and Fig.~\ref{fig:MR_Hallangle_Sub_GeMn_HB} show a strong similarity to the underlying substrate. This suggests that neither the inhomogeneity, magnetic nature nor the presence of Mn acceptors in the GeMn thin film leads to the observed transport properties. We conclude that the transport properties of the GeMn sample do emerge from parallel conduction through the substrate. 

This dominating contribution of the substrate can be understood, when one considers the system of an epitaxially fabricated GeMn thin film on top of the high resistivity Ge substrate as two parallel conducting resistors. For an independent determination of the transport properties of the epi-layer without contributions from the substrate layer, the resistance of the GeMn epi-layer has to be at least a factor of ten smaller than that of the substrate. A comparison of the thicknesses of these two layers implicates that the epi-layer resistivity then has to be smaller by a factor of $10^4$ than the resistivity of the Ge substrate. Considering the RT value of the substrate resistivity of about \SI{40}{\ohm\centi\meter} this in turn means that the GeMn epi-layer resistivity has to be in the \SI{e-3}{\ohm\centi\metre} regime. For the present GeMn epitaxial layer, having a hole density around \SI{e19}{\per\cubic\centi\metre} (cf. section~\ref{sec:nSub}), but nevertheless being  non-degenerate, that would demand RT mobilities in the order of a few \SI{e2}{\centi\metre\squared\per\volt\per\second}. However, since GeMn thin films exhibit a very inhomogeneous nanostructure,\cite{Bou2006,Bougeard2009,Jam2006,Li2007} such a high mobility can not be expected. In fact, mobilities of that order of magnitude are only reached in conventional p-type doped Ge with similar hole concentrations, when the dopants are homogeneously diluted in the host matrix.\cite{Golikova1962} In essence, because of the low conductivity of non-degenerate GeMn thin films, one can not determine the transport properties of the GeMn epi-layer in a straightforward manner, when it is fabricated on high resistivity Ge substrates.

\subsection{Degenerate epitaxial p-type Ge on high resistivity Ge substrates\label{sec:twolayer}}
\begin{figure}
	\centering
		\includegraphics{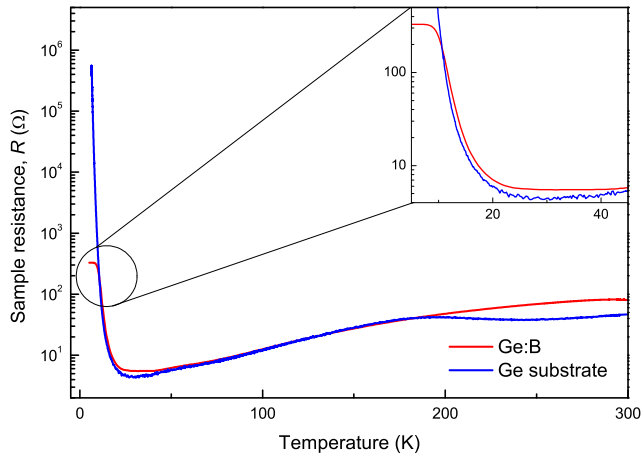}
	   \caption{Sample resistance as function of temperature for the Ge:B sample with a doping concentration of  \SI{5e19}{\per\cubic\centi\metre}. The reference measurement of the substrate is also depicted. The inset shows a close-up for small temperatures, indicating the metallic conductance of the Ge:B sample, when parallel conduction through the substrate ceases.\label{fig:rho(T)_GeB}}
\end{figure}
We now would like to address whether an electrical transport characterization of degenerately doped, p-type GeMn thin films on high resistivity Ge substrates, i.e. thin films with carrier concentrations clearly above \SI{e19}{\per\cubic\centi\metre}, is feasible. In order to separate phenomena related to the magnetization or nanostructure from those related to parasitic conduction through the substrate, we explored degenerately doped, non-magnetic Ge:B epi-layers as a model system. Since B opposed to Mn, is not a deep, but a shallow acceptor in Ge, a doping concentration of \SI{5e19}{\per\cubic\centi\metre} lies well above the Mott insulator-to-metal transition, and is therefore sufficiently large to deliver thin films with degenerate, metallic-like conduction properties. Figure~\ref{fig:rho(T)_GeB} shows the temperature dependence of the resistance of the Ge:B sample. Interestingly, only at temperatures below approximately \SI{10}{\kelvin} the measurement reflects the metallic character of the epi-layer, as the resistance enters a constant value regime. Above this temperature the curve quickly traces the measurement of the Ge substrate, which is also depicted for a comparison.
\begin{figure}
	\centering
		\includegraphics{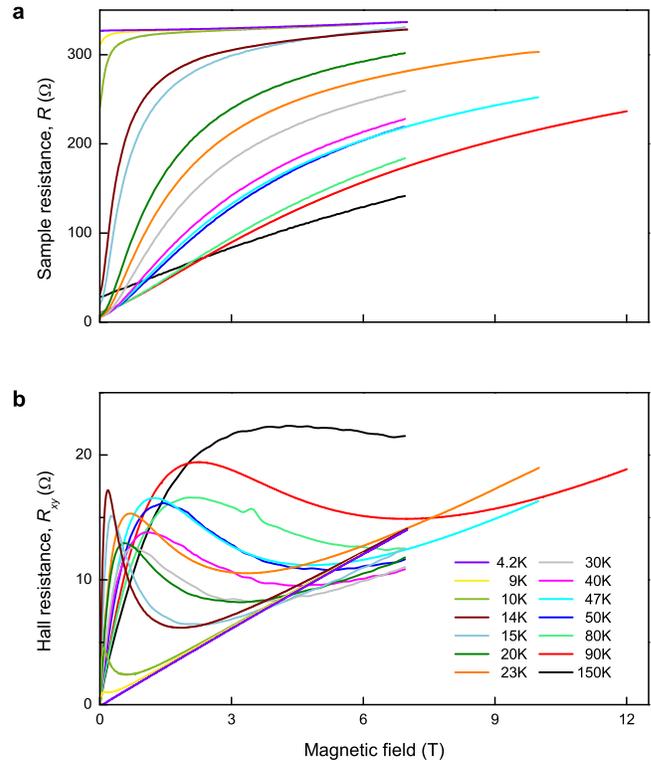}
	   \caption{\textbf{(a)} Sample resistance versus magnetic field of the Ge:B sample. \textbf{(b)} Hall resistance of the Ge:B sample. The dip of $R$ at zero field and the peak of $R_{xy}$ at low fields mark the onset of parallel conduction through the substrate at \SI{9}{\kelvin}. The color code is the same in both panels.\label{fig:MR_Hall_GeB_measurement}}
\end{figure}

The magnetic field dependence of the longitudinal resistance of the Ge:B sample is depicted in Figure~\hyperref[fig:MR_Hall_GeB_measurement]{\ref{fig:MR_Hall_GeB_measurement}(a)} for different temperatures. At temperatures below 
\SI{9}{\kelvin}, the resistance shows little field dependence, yielding a MR effect which does not exceed \SI{3}{\percent} at \SI{4.2}{\kelvin} and \SI{7}{\tesla}. At approximately \SI{9}{\kelvin} we observe the onset of the decrease of the zero field resistance with increasing temperature, as already depicted in Fig.~\ref{fig:rho(T)_GeB}. However, the resistance now rises quickly with increasing magnetic field, hence giving an increased MR effect. Eventually at high fields the resistance tends to saturate at the \SI{4.2}{\kelvin} value. With increasing measurement temperature this saturation is shifted towards higher fields, while at the highest temperatures full saturation is not reached anymore within the investigated field range. 

The Hall effect of the Ge:B sample is shown in Fig.~\hyperref[fig:MR_Hall_GeB_measurement]{\ref{fig:MR_Hall_GeB_measurement}(b)}. Below \SI{9}{\kelvin} we observe a linear Hall effect. At temperatures of \SI{9}{\kelvin} and above the field dependence drastically changes: For small field values we first observe a strong increase of the Hall slope. Upon increasing the field, the Hall effect shows a peak-like maximum and then approaches the Hall curve measured for \SI{4.2}{\kelvin} asymptotically.

The transport behavior of the Ge:B film, depicted in Fig.~\ref{fig:rho(T)_GeB} and Fig.~\hyperref[fig:MR_Hall_GeB_measurement]{\ref{fig:MR_Hall_GeB_measurement}} is not in line with the metallic character of the epi-layer. We can rather identify two distinct temperature regimes with different properties below and above \SI{9}{\kelvin}. Similar results for the Hall effect and field dependence of longitudinal resistivity which can be separated into two temperature regimes were found for the GeMn material systems in degenerate thin films prepared by ion implantation.\cite{Riss2009} They were interpreted in terms of a two-band like conduction scheme, accounting for possible electronic ground and excited states of Mn in Ge.\cite{Zhou2009} In contrast, our results in the two distinct regimes are naturally explained by assuming parallel conduction through the substrate: Below \SI{9}{\kelvin} parallel conduction is not present, since the substrate resistance gets very large, whereas the resistance of the metallic epi-layer does not change. Assuming now that conduction only takes place in the \SI{200}{\nano\metre} thick Ge:B epi-layer, we can extract a hole density of \SI{1.54e19}{\per\cubic\centi\metre}, in fair agreement with the nominal concentration value. Furthermore the carrier mobility amounts to the relatively small value of $\mu =  \SI{310}{\centi\metre\squared\per\volt\per\second}$. Due to the general proportionality between the orbital MR and carrier mobility, the small MR effect would therefore also be in line with conduction through the metallic Ge:B epi-layer. The transport measurements of the Ge:B sample can undoubtedly be attributed solely to the Ge:B epi-layer in the temperature regime below \SI{9}{\kelvin}. For temperatures above \SI{9}{\kelvin} we need to include substrate contributions for an interpretation of the magneto transport data. At low fields, conduction will mostly take place in the substrate because of its smaller resistance. At large fields, conduction through the substrate will quickly cease, because its MR gets larger. Then most of the current flows through the epi-layer. Thus, in the Hall as well as the MR measurement, we probe the substrate properties at small magnetic field and the Ge:B epi-layer properties at high field, leading to the described peaking and saturation effects. The decreasing tendency of saturation at high magnetic field with increasing measurement temperature comes from the weakening of MR of the Ge substrate. The Ge:B sample thus has to be regarded as a system of two conducting layers with different galvanomagnetic responses.

This phenomenological interpretation is supported by a description of the magneto transport data with a two-layer conduction model. It is based on the assumption that each conducting layer can be described by its individual resistivity tensor, which reduces to a $2\times2$ matrix in the case where the magnetic field is normal to the plane of carrier motion. Since the two layers are not equally thick, in the following expressions we will give sheet resistivities rather than bulk resistivities to maintain generality. \footnote{In the case of a two-carrier-type model the same equations apply, when the sheet terms are replaced by their respective bulk counterparts.} The resulting components of the sheet resistivity tensor of the combined two-layer system have then the form
\begin{widetext}
\begin{eqnarray}
\rho_{xx} &	=  & \rho_{yy}  =  \frac{\rho_{\mathrm{1,}xx}({\rho_{\mathrm{2,}xx}}^2 + {\rho_{\mathrm{2,}xy}}^2) + 	\rho_{\mathrm{2,}xx}({\rho_{\mathrm{1,}xx}}^2 + {\rho_{\mathrm{1,}xy}}^2)}{(\rho_{\mathrm{1,}xx}+\rho_{\mathrm{2,}xx})^2+
(\rho_{\mathrm{1,}xy}+\rho_{\mathrm{2,}xy})^2} \nonumber\\
\rho_{xy}	& = -& \rho_{yx}  =  \frac{\rho_{\mathrm{1,}xy}({\rho_{\mathrm{2,}xx}}^2 + {\rho_{\mathrm{2,}xy}}^2) + 	\rho_{\mathrm{2,}xy}({\rho_{\mathrm{1,}xx}}^2 + {\rho_{\mathrm{1,}xy}}^2)}{(\rho_{\mathrm{1,}xx}+\rho_{\mathrm{2,}xx})^2+
(\rho_{\mathrm{1,}xy}+\rho_{\mathrm{2,}xy})^2}.
\label{eq:resistivity_tensor}
\end{eqnarray}
\end{widetext}
The quantities with subscripts 1 and 2 correspond to the tensor components of the Ge:B epi-layer and the Ge substrate layer, respectively. We restrict ourselves to a semi-empirical application of the above equations for a computation of the Hall and MR effect of the Ge:B sample for different temperatures. The parameters $\rho_{1/2,xx}$ and $\rho_{1/2,xy}$ entering this computation are taken from measurements: The contributions $\rho_{\mathrm{1,}xx}$ and $\rho_{\mathrm{1,}xy}$ of Ge:B epi-layer correspond to the \SI{4.2}{\kelvin} measurement curves of the Ge:B sample as were shown in Fig.~\hyperref[fig:MR_Hall_GeB_measurement]{\ref{fig:MR_Hall_GeB_measurement}(a)} and \hyperref[fig:MR_Hall_GeB_measurement]{(b)}. We assume they do not vary with increasing temperature, which is justified by the metallic character of this epi-layer (see also section~\ref{sec:nSub}) therefore using them for all temperatures we investigate. The contributions of the Ge substrate, $\rho_{\mathrm{2,}xx}$ and $\rho_{\mathrm{2,}xy}$, are taken from the measurements depicted in Fig.~\hyperref[fig:MR_Hallangle_Sub_GeMn_HB]{\ref{fig:MR_Hallangle_Sub_GeMn_HB}(b)} and \hyperref[fig:MR_Hallangle_Sub_GeMn_HB]{(c)} for each corresponding temperature. 
\begin{figure}
	\includegraphics{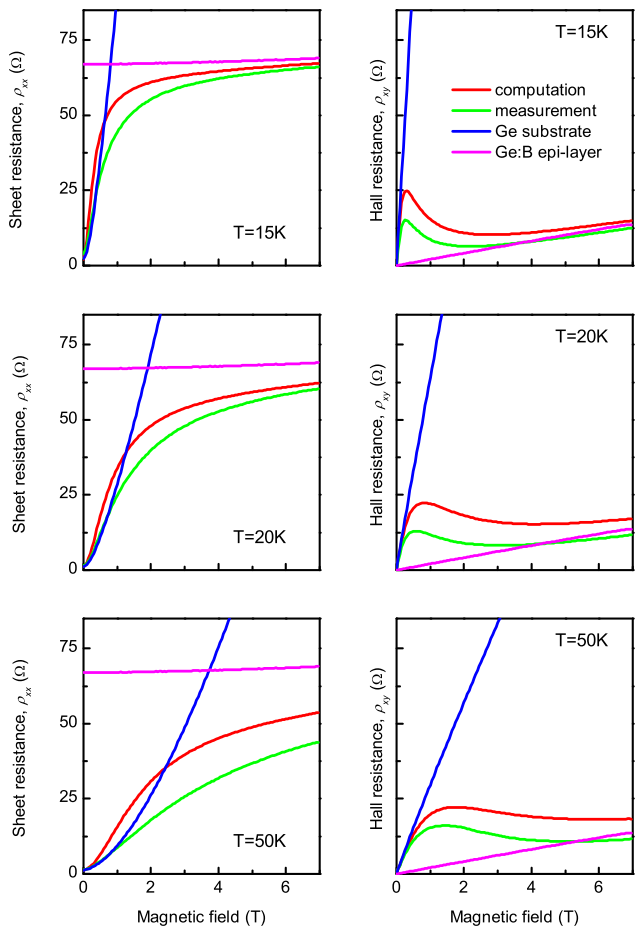}
	 \caption{Sheet (left) and Hall resistance (right) versus magnetic field for different temperatures of the Ge:B sample affected by parallel conduction. Shown is the measurement, the computation according to Eqs.~(\ref{eq:resistivity_tensor}) and the individual contributions of the substrate and metallic epi-layer.\label{fig:MR_Hall_GeB_computation}}	
\end{figure}

Figure~\ref{fig:MR_Hall_GeB_computation} shows in the left and right panels the results of the computation of $\rho_{xx}$ and $\rho_{xy}$, respectively, compared to the measured values for \SI{15}{\kelvin}, \SI{20}{\kelvin} and  \SI{50}{\kelvin}. We also included the $\rho_{1/2,xx}$ and $\rho_{1/2,xy}$ contributions of the Ge:B epi-layer and the substrate in the plot.  There is a good agreement between the two-layer conduction model and the measurements of the Ge:B sample for both, the $\rho_{xx}$ and $\rho_{xy}$ component. In particular the low field domination of the Ge substrate layer as well as the saturation for high field at the Ge:B epi-layer contribution can be reproduced well by Eqs.~\eqref{eq:resistivity_tensor}. Thus, the model qualitatively demonstrates that parallel conduction through the substrate is also present in a sample with a degenerate, metallic, well conducting epi-layer in the extrinsic range of the underlying substrate.
For the small quantitative differences between the computed and the actual experimental results, a major reason can be made out. The model Eqs.~\eqref{eq:resistivity_tensor} strictly apply for a two-layer system, where both layers have the same in-plane geometry. Since the Hall bar mesa, however, does not define such a geometry for the substrate conduction channel, differences between theory and experiment will occur. 

From the results on the degenerate Ge:B reference sample we infer that in the case of metallic GeMn thin films deposited on the high resistivity substrate intrinsic properties of the GeMn epi-layer may be directly derived in the freeze-out temperature regime of the substrate. For higher temperatures, however, care must be taken to separate the intrinsic properties of GeMn from the aforementioned effects arising due to the two-layer conduction. 

\subsection{Influence of the sample geometry on MR measurements\label{sec:vdPgeometry}}
\begin{figure}
	\includegraphics{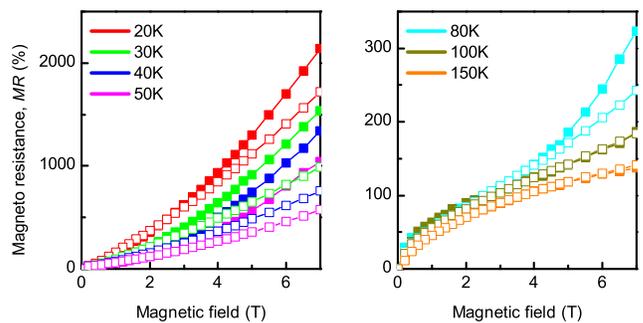}
	 \caption{Transversal MR for various temperatures for the GeMn sample (open symbols) and the bare Ge substrate (filled symbols).\label{fig:MR_Sub_GeMn_vdP}}	
\end{figure}

It was previously shown in Fig.~\hyperref[fig:MR_Hallangle_Sub_GeMn_HB]{\ref{fig:MR_Hallangle_Sub_GeMn_HB}(b)} that the high resistivity substrate of very pure Ge exhibits an extremely large MR effect up to \SI{50000}{\percent}. However, reports of magneto transport properties of high purity Ge show that this large effect is not expected.\cite{Goldberg1956,Gallagher1967} To study the MR effect of the high resistivity substrate further, we fabricated additional samples using a van der Pauw (vdP) geometry, as depicted in the schematic of Fig.~\hyperref[fig:HB_vdP]{\ref{fig:HB_vdP}(b)}. Interestingly, the MR effect of the vdP sample presented in Figure~\ref{fig:MR_Sub_GeMn_vdP} is now more than twenty times smaller than for the corresponding Hall bar sample. Evidently, the large MR effect previously obtained in Hall bar geometry is not an inherent physical property of the Ge substrate. The measurements taken in vdP geometry agree much better with the above mentioned magneto-transport studies of Ge.\cite{Gallagher1967,Goldberg1956}

Up to now we can only speculate about the reasons inducing the large MR effect. Most probably it is related to a redistribution of the current lines upon applying a magnetic field, similar to the effect observed in ref.~\citenum{Solin2000}. Referring to the sample schematic depicted in Fig.~\hyperref[fig:HB_vdP]{\ref{fig:HB_vdP}(a)} the relatively large, metalized voltage probes 2, 3, 4, 6, 7 and 8 have to be considered as effective short-circuits within the sample volume, if conduction through substrate is present. In a magnetic field these short-circuits may be diminished as the electric field and therefore the current flow will get tangent to these areas. The result is an artificial increase of the sample resistance. Additionally the small length to width ratio of $l/w\approx \mathrm{2}$, basically given by the separation of contacts 1 and 5 and the width of contact 1, may also favor a considerable Hall effect induced contribution to the intrinsic MR.\cite{Wick1954,Jensen1972} 

We also reinvestigated the GeMn thin film sample in vdP geometry. Upon comparing it with the Ge substrate in Fig.~\ref{fig:MR_Sub_GeMn_vdP} we now remark a pronounced difference between the MR measurement above approximately \SI{3}{\tesla} for temperatures up to \SI{80}{\kelvin}. The MR of the GeMn sample follows a more linear behavior, while the MR of the substrate still increases superlinearly above this field. At \SI{100}{\kelvin} and higher temperatures both samples exhibit the same MR effect. We infer from this behavior that at least below \SI{100}{\kelvin} transport properties of the GeMn epi-layer get visible in the measurement. This may be due to the fact, that in vdP geometry the volume of the epi-layer is not restricted to the in-plane dimensions of the etched Hall bar mesa, but extends over the whole chip area. Therefore the effective volume ratio of the epi-layer to the substrate, hence the conductance ratio, is increased in vdP geometry compared to the Hall bar geometry. 

We thus conclude that a vdP geometry is to be favored over a Hall bar in the GeMn material system with its high probability of parallel conduction through the substrate. By using a vdP geometry the comparison of experimental data with an elaborate, ab-initio two-layer conduction model, extending the scheme outlined in section~\ref{sec:twolayer}, may enable a derivation of inherent transport properties of GeMn thin films, in spite of the dominant contribution of the Ge substrate.

\subsection{GeMn on n-type Ge substrates\label{sec:nSub}}
\begin{figure}
	\centering
		\includegraphics{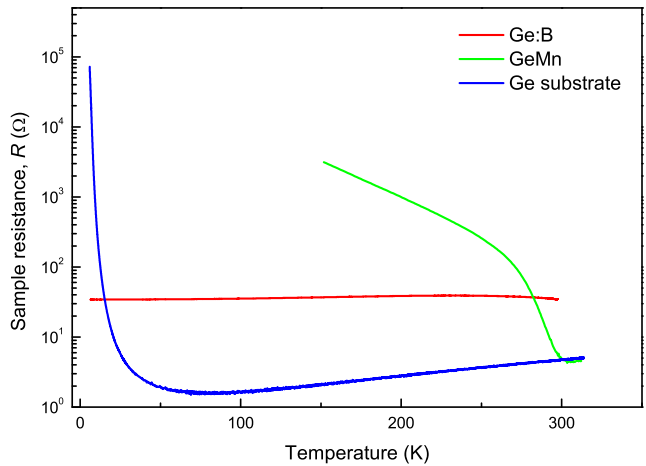}
	   \caption{Sample resistance versus temperature for the GeMn sample (green), the Ge:B sample grown on the n-type Ge substrate (blue). The GeMn sample is affected by parallel conduction above \SI{250}{\kelvin} as the pn-barrier gets inefficient. The Ge:B sample shows metallic conduction up to RT.	\label{fig:rho(T)_GeMn_GeB_nGesub}}
\end{figure}
The derivation of the GeMn transport properties would be much easier, if the substrate contribution could be further reduced. To this end we fabricated another GeMn sample, which employs a Ge substrate with RT resistivity of \SI{0.13}{\ohm\centi\metre} having a well defined concentration of Sb donors. The build up of a rectifying interface between the p-type GeMn epi-layer and the n-type substrate would isolate the epi-layer electrically from the substrate. The \SI{80}{\nano\meter} thick GeMn epi-layer has a Mn concentration of \SI{10}{\percent}. To test the benefit of this concept, we again deposited a metallic Ge:B epi-layer on such a substrate. Transport measurements of these samples were made in vdP geometry.

Figure~\ref{fig:rho(T)_GeMn_GeB_nGesub} displays the temperature dependent resistance of the second Ge:B sample together with the employed substrate.  The curve shape is now in agreement with the metallic character of the Ge:B thin film over the entire temperature range and is clearly different from the substrate behavior. Note that the substrate resistance is actually a factor of ten smaller than the resistance of the epi-layer for temperatures above \SI{30}{\kelvin}, demonstrating the effectivity of the rectifying pn-barrier. The absence of parallel conduction through the substrate is also reflected in Hall measurements (not shown), which in contrast to the measurements depicted in Fig.~\ref{fig:MR_Hall_GeB_measurement} exhibit a linear Hall effect up to RT corresponding to the nominal doping concentration of the thin film. 

Also shown in Fig.~\ref{fig:rho(T)_GeMn_GeB_nGesub} is then the resistance curve of the GeMn sample grown on the n-type substrate. Reliable measurement data are only available above \SI{150}{\kelvin}, since the electrical contacts become non-ohmic below this temperature. Nevertheless, in the available temperature range we remark a clear difference of the resistance of this sample compared to the substrate. The resistance becomes quickly larger, suggesting the absence of parallel conduction through the substrate. Magneto transport studies of this sample reveal p-type conduction pointing towards the acceptor role of Mn in Ge. Hole concentrations ranging from \SI{5e18}{\per\cubic\centi\metre} to \SI{1.5e19}{\per\cubic\centi\metre} for temperatures between \SI{150}{\kelvin} and \SI{210}{\kelvin} could be deduced from the high field slope of Hall measurements (not shown). We could not identify any signs of a magnetization induced AHE which, in general pointing towards a low polarization of the holes, could also be related to the decreasing magnetic response of these types of GeMn thin films at the accessible temperatures above \SI{150}{\kelvin}.\cite{Bougeard2009, Bou2006} 

\begin{figure}
	\includegraphics{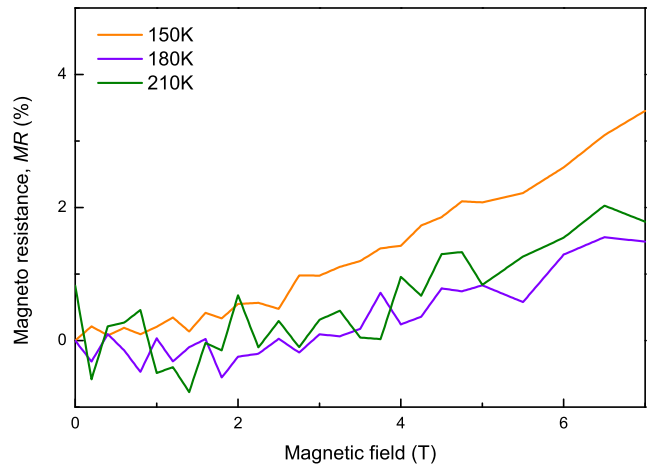}
	 \caption{Transversal MR for various temperatures for the GeMn sample grown on the n-type Ge substrate.\label{fig:MR_nSub_GeMn_vdP}}	
\end{figure}
Figure~\ref{fig:MR_nSub_GeMn_vdP} depicts the MR of the GeMn sample for three different temperatures. Interestingly, the MR changes only slightly with temperature and does not exceed \SI{3}{\percent} for the highest field, in contrast to a value of \SI{120}{\percent} for the sample affected by parallel conduction through the substrate, as depicted in Fig.~\ref{fig:MR_Sub_GeMn_vdP}. While this underlines the immense influence of the substrate contribution, it also demonstrates that the MR in our GeMn samples is apparently rather small. Its order of magnitude as well as the parabolic curvature could very well originate from the ubiquitous, normal orbital MR in a low mobility conductor.

The usage of an n-type substrate seems to unveil the intrinsic transport characteristics of our GeMn thin film sample. Parallel conduction through the substrate is greatly reduced, compared to thin films fabricated on the high resistivity Ge substrate. For degenerate epitaxial GeMn thin films a slightly n-type doped Ge substrate seems to be most adequate for transport studies. 
However, there are several drawbacks from a general, straightforward application of this approach to non-degenerate samples: The pn-barrier concept fails for higher temperatures, as indicated by the large drop of the resistance curve of the GeMn sample in Fig.~\ref{fig:rho(T)_GeMn_GeB_nGesub} above \SI{250}{\kelvin}. A RT characterization of such devices is impossible, as parallel conduction through the substrate will be present. Moreover, a rapid thermal annealing process to obtain ohmic contacts could not be used for these type of samples, as this resulted in a direct contact to the substrate which short circuits the pn-barrier. A laser assisted ultra short time annealing employed instead, however, did not provide ohmic contacts that work down to cryogenic temperatures. More sophisticated approaches might be explored to overcome this limitation and enable also low temperature measurements.

\section{Summary}
In summary, the work presented here has shown that transport phenomena of non-degenerate GeMn thin films with hole concentrations around \SI{e19}{\per\cubic\centi\metre} fabricated on high resistivity Ge substrates are not a consequence of the alloying of Mn with Ge. Instead it was found that the transport studies are severely influenced by parallel conduction through the substrate. This is in essence due to comparable resistances of the individual conducting layers. In this context findings of extremely large magneto resistance effects up to \SI{50000}{\percent} are related to an unfavorable measurement geometry. Measurements of a degenerate, p-type Ge:B reference sample showed that despite of the well conducting epi-layer, parallel conduction through the substrate is also present, significantly altering Hall and MR measurements which could only be understood in a two-layer conduction model. Parallel conduction through the substrate has been fully suppressed for the degenerate, p-type reference and partly for the GeMn thin film sample, by using Sb doped n-type substrates. 

Our results hint towards the importance of a thorough characterization of the substrate properties in transport studies of GeMn thin films that are fabricated on Ge substrates. An increasing awareness and proper understanding of this problem may help to rule out possible misinterpretations. Such misinterpretation may a priori be avoided by using semi-insulating GaAs substrates with a resistivity larger than \SI{e3}{\ohm\centi\metre}, delivering a small lattice mismatch to the Ge diamond lattice structure. A few reports on magneto transport of various types of GeMn thin films employing GaAs substrates exist,\cite{Par2002,Li2006a,Zen2006,Zeng2008} however, out-diffusion of As or Ga is a critical issue.\cite{Tsuchida2009,Par2001} It was recently shown that this type of unintentional co-doping may actually change the structural and also magnetic properties of GeMn thin films.\cite{Yu2010} 

\begin{acknowledgments}
The authors would like to thank the Deutsche Forschungsgemeinschaft for financial support via SPP 1285.
\end{acknowledgments}


\begin{thebibliography}{32}%
\makeatletter
\providecommand \@ifxundefined [1]{%
 \@ifx{#1\undefined}
}%
\providecommand \@ifnum [1]{%
 \ifnum #1\expandafter \@firstoftwo
 \else \expandafter \@secondoftwo
 \fi
}%
\providecommand \@ifx [1]{%
 \ifx #1\expandafter \@firstoftwo
 \else \expandafter \@secondoftwo
 \fi
}%
\providecommand \natexlab [1]{#1}%
\providecommand \enquote  [1]{``#1''}%
\providecommand \bibnamefont  [1]{#1}%
\providecommand \bibfnamefont [1]{#1}%
\providecommand \citenamefont [1]{#1}%
\providecommand \href@noop [0]{\@secondoftwo}%
\providecommand \href [0]{\begingroup \@sanitize@url \@href}%
\providecommand \@href[1]{\@@startlink{#1}\@@href}%
\providecommand \@@href[1]{\endgroup#1\@@endlink}%
\providecommand \@sanitize@url [0]{\catcode `\\12\catcode `\$12\catcode
  `\&12\catcode `\#12\catcode `\^12\catcode `\_12\catcode `\%12\relax}%
\providecommand \@@startlink[1]{}%
\providecommand \@@endlink[0]{}%
\providecommand \url  [0]{\begingroup\@sanitize@url \@url }%
\providecommand \@url [1]{\endgroup\@href {#1}{\urlprefix }}%
\providecommand \urlprefix  [0]{URL }%
\providecommand \Eprint [0]{\href }%
\providecommand \doibase [0]{http://dx.doi.org/}%
\providecommand \selectlanguage [0]{\@gobble}%
\providecommand \bibinfo  [0]{\@secondoftwo}%
\providecommand \bibfield  [0]{\@secondoftwo}%
\providecommand \translation [1]{[#1]}%
\providecommand \BibitemOpen [0]{}%
\providecommand \bibitemStop [0]{}%
\providecommand \bibitemNoStop [0]{.\EOS\space}%
\providecommand \EOS [0]{\spacefactor3000\relax}%
\providecommand \BibitemShut  [1]{\csname bibitem#1\endcsname}%
\let\auto@bib@innerbib\@empty
%</preamble>
\bibitem [{\citenamefont {Bougeard}\ \emph {et~al.}(2006)\citenamefont
  {Bougeard}, \citenamefont {Ahlers}, \citenamefont {Trampert}, \citenamefont
  {Sircar},\ and\ \citenamefont {Abstreiter}}]{Bou2006}%
  \BibitemOpen
  \bibfield  {author} {\bibinfo {author} {\bibfnamefont {D.}~\bibnamefont
  {Bougeard}}, \bibinfo {author} {\bibfnamefont {S.}~\bibnamefont {Ahlers}},
  \bibinfo {author} {\bibfnamefont {A.}~\bibnamefont {Trampert}}, \bibinfo
  {author} {\bibfnamefont {N.}~\bibnamefont {Sircar}}, \ and\ \bibinfo {author}
  {\bibfnamefont {G.}~\bibnamefont {Abstreiter}},\ }\href {\doibase 10.1103/PhysRevLett.97.237202}
  {\bibfield  {journal} {\bibinfo  {journal}
  {Phys. Rev. Lett.}\ }\textbf {\bibinfo {volume} {97}},\ \bibinfo {eid}
  {237202} (\bibinfo {year} {2006})}\BibitemShut {NoStop}%
\bibitem [{\citenamefont {Jamet}\ \emph {et~al.}(2006)\citenamefont {Jamet},
  \citenamefont {Barski}, \citenamefont {Devillers}, \citenamefont {Poydenot},
  \citenamefont {Dujardin}, \citenamefont {Bayle-Guillemaud}, \citenamefont
  {Rothman}, \citenamefont {Bellet-Amalric}, \citenamefont {Marty},
  \citenamefont {Cibert}, \citenamefont {Mattana},\ and\ \citenamefont
  {Tatarenko}}]{Jam2006}%
  \BibitemOpen
  \bibfield  {author} {\bibinfo {author} {\bibfnamefont {M.}~\bibnamefont
  {Jamet}}, \bibinfo {author} {\bibfnamefont {A.}~\bibnamefont {Barski}},
  \bibinfo {author} {\bibfnamefont {T.}~\bibnamefont {Devillers}}, \bibinfo
  {author} {\bibfnamefont {V.}~\bibnamefont {Poydenot}}, \bibinfo {author}
  {\bibfnamefont {R.}~\bibnamefont {Dujardin}}, \bibinfo {author}
  {\bibfnamefont {P.}~\bibnamefont {Bayle-Guillemaud}}, \bibinfo {author}
  {\bibfnamefont {J.}~\bibnamefont {Rothman}}, \bibinfo {author} {\bibfnamefont
  {E.}~\bibnamefont {Bellet-Amalric}}, \bibinfo {author} {\bibfnamefont
  {A.}~\bibnamefont {Marty}}, \bibinfo {author} {\bibfnamefont
  {J.}~\bibnamefont {Cibert}}, \bibinfo {author} {\bibfnamefont
  {R.}~\bibnamefont {Mattana}}, \ and\ \bibinfo {author} {\bibfnamefont
  {S.}~\bibnamefont {Tatarenko}},\ }\href {\doibase 10.1038/nmat1686}
  {\bibfield  {journal} {\bibinfo  {journal} {Nature Materials}\ }\textbf
  {\bibinfo {volume} {5}},\ \bibinfo {pages} {653} (\bibinfo {year}
  {2006})}\BibitemShut {NoStop}%
\bibitem [{\citenamefont {Bougeard}\ \emph {et~al.}(2009)\citenamefont
  {Bougeard}, \citenamefont {Sircar}, \citenamefont {Ahlers}, \citenamefont
  {Lang}, \citenamefont {Abstreiter}, \citenamefont {Trampert}, \citenamefont
  {LeBeau}, \citenamefont {Stemmer}, \citenamefont {Saxey},\ and\ \citenamefont
  {Cerezo}}]{Bougeard2009}%
  \BibitemOpen
  \bibfield  {author} {\bibinfo {author} {\bibfnamefont {D.}~\bibnamefont
  {Bougeard}}, \bibinfo {author} {\bibfnamefont {N.}~\bibnamefont {Sircar}},
  \bibinfo {author} {\bibfnamefont {S.}~\bibnamefont {Ahlers}}, \bibinfo
  {author} {\bibfnamefont {V.}~\bibnamefont {Lang}}, \bibinfo {author}
  {\bibfnamefont {G.}~\bibnamefont {Abstreiter}}, \bibinfo {author}
  {\bibfnamefont {A.}~\bibnamefont {Trampert}}, \bibinfo {author}
  {\bibfnamefont {J.~M.}\ \bibnamefont {LeBeau}}, \bibinfo {author}
  {\bibfnamefont {S.}~\bibnamefont {Stemmer}}, \bibinfo {author} {\bibfnamefont
  {D.~W.}\ \bibnamefont {Saxey}}, \ and\ \bibinfo {author} {\bibfnamefont
  {A.}~\bibnamefont {Cerezo}},\ }\href {\doibase 10.1021/nl901928f} {\bibfield
  {journal} {\bibinfo  {journal} {Nano Letters}\ }\textbf {\bibinfo {volume}
  {9}},\ \bibinfo {pages} {3743} (\bibinfo {year} {2009})}\BibitemShut
  {NoStop}%
\bibitem [{\citenamefont {Li}\ \emph {et~al.}(2007)\citenamefont {Li},
  \citenamefont {Zeng}, \citenamefont {van Benthem}, \citenamefont {Chisholm},
  \citenamefont {Shen}, \citenamefont {Rao}, \citenamefont {Dixit},
  \citenamefont {Feldman}, \citenamefont {Petukhov}, \citenamefont {Foygel},\
  and\ \citenamefont {Weitering}}]{Li2007}%
  \BibitemOpen
  \bibfield  {author} {\bibinfo {author} {\bibfnamefont {A.~P.}\ \bibnamefont
  {Li}}, \bibinfo {author} {\bibfnamefont {C.}~\bibnamefont {Zeng}}, \bibinfo
  {author} {\bibfnamefont {K.}~\bibnamefont {van Benthem}}, \bibinfo {author}
  {\bibfnamefont {M.~F.}\ \bibnamefont {Chisholm}}, \bibinfo {author}
  {\bibfnamefont {J.}~\bibnamefont {Shen}}, \bibinfo {author} {\bibfnamefont
  {S.~V. S.~N.}\ \bibnamefont {Rao}}, \bibinfo {author} {\bibfnamefont {S.~K.}\
  \bibnamefont {Dixit}}, \bibinfo {author} {\bibfnamefont {L.~C.}\ \bibnamefont
  {Feldman}}, \bibinfo {author} {\bibfnamefont {A.~G.}\ \bibnamefont
  {Petukhov}}, \bibinfo {author} {\bibfnamefont {M.}~\bibnamefont {Foygel}}, \
  and\ \bibinfo {author} {\bibfnamefont {H.~H.}\ \bibnamefont {Weitering}},\
  }\href {\doibase 10.1103/PhysRevB.75.201201} {\bibfield  {journal} {\bibinfo
  {journal} {Phys. Rev. B.}\ }\textbf {\bibinfo {volume} {75}},\ \bibinfo {eid}
  {201201} (\bibinfo {year} {2007})}\BibitemShut {NoStop}%
\bibitem [{\citenamefont {Park}\ \emph {et~al.}(2002)\citenamefont {Park},
  \citenamefont {Hanbicki}, \citenamefont {Erwin}, \citenamefont {Hellberg},
  \citenamefont {Sullivan}, \citenamefont {Mattson}, \citenamefont {Ambrose},
  \citenamefont {Wilson}, \citenamefont {Spanos},\ and\ \citenamefont
  {Jonker}}]{Par2002}%
  \BibitemOpen
  \bibfield  {author} {\bibinfo {author} {\bibfnamefont {Y.~D.}\ \bibnamefont
  {Park}}, \bibinfo {author} {\bibfnamefont {A.~T.}\ \bibnamefont {Hanbicki}},
  \bibinfo {author} {\bibfnamefont {S.~C.}\ \bibnamefont {Erwin}}, \bibinfo
  {author} {\bibfnamefont {C.~S.}\ \bibnamefont {Hellberg}}, \bibinfo {author}
  {\bibfnamefont {J.~M.}\ \bibnamefont {Sullivan}}, \bibinfo {author}
  {\bibfnamefont {J.~E.}\ \bibnamefont {Mattson}}, \bibinfo {author}
  {\bibfnamefont {T.~F.}\ \bibnamefont {Ambrose}}, \bibinfo {author}
  {\bibfnamefont {A.}~\bibnamefont {Wilson}}, \bibinfo {author} {\bibfnamefont
  {G.}~\bibnamefont {Spanos}}, \ and\ \bibinfo {author} {\bibfnamefont {B.~T.}\
  \bibnamefont {Jonker}},\ }\href {\doibase 10.1126/science.1066348} {\bibfield
   {journal} {\bibinfo  {journal} {Science}\ }\textbf {\bibinfo {volume}
  {295}},\ \bibinfo {pages} {651} (\bibinfo {year} {2002})}\BibitemShut
  {NoStop}%
\bibitem [{\citenamefont {Li}\ \emph {et~al.}(2005)\citenamefont {Li},
  \citenamefont {Wendelken}, \citenamefont {Shen}, \citenamefont {Feldman},
  \citenamefont {Thompson},\ and\ \citenamefont {Weitering}}]{Li2005a}%
  \BibitemOpen
  \bibfield  {author} {\bibinfo {author} {\bibfnamefont {A.~P.}\ \bibnamefont
  {Li}}, \bibinfo {author} {\bibfnamefont {J.~F.}\ \bibnamefont {Wendelken}},
  \bibinfo {author} {\bibfnamefont {J.}~\bibnamefont {Shen}}, \bibinfo {author}
  {\bibfnamefont {L.~C.}\ \bibnamefont {Feldman}}, \bibinfo {author}
  {\bibfnamefont {J.~R.}\ \bibnamefont {Thompson}}, \ and\ \bibinfo {author}
  {\bibfnamefont {H.~H.}\ \bibnamefont {Weitering}},\ }\href {\doibase 10.1103/PhysRevB.72.195205}
  {\bibfield  {journal} {\bibinfo  {journal} {Phys.
  Rev. B.}\ }\textbf {\bibinfo {volume} {72}},\ \bibinfo {pages} {195205}
  (\bibinfo {year} {2005})}\BibitemShut {NoStop}%
\bibitem [{\citenamefont {Zhou}\ \emph {et~al.}(2010)\citenamefont {Zhou},
  \citenamefont {B\"urger}, \citenamefont {M\"ucklich}, \citenamefont {Baumgart},
  \citenamefont {Skorupa}, \citenamefont {Timm}, \citenamefont {Oesterlin},
  \citenamefont {Helm},\ and\ \citenamefont {Schmidt}}]{Zhou2010}%
  \BibitemOpen
  \bibfield  {author} {\bibinfo {author} {\bibfnamefont {S.}~\bibnamefont
  {Zhou}}, \bibinfo {author} {\bibfnamefont {D.}~\bibnamefont {B\"urger}},
  \bibinfo {author} {\bibfnamefont {A.}~\bibnamefont {M\"ucklich}}, \bibinfo
  {author} {\bibfnamefont {C.}~\bibnamefont {Baumgart}}, \bibinfo {author}
  {\bibfnamefont {W.}~\bibnamefont {Skorupa}}, \bibinfo {author} {\bibfnamefont
  {C.}~\bibnamefont {Timm}}, \bibinfo {author} {\bibfnamefont {P.}~\bibnamefont
  {Oesterlin}}, \bibinfo {author} {\bibfnamefont {M.}~\bibnamefont {Helm}}, \
  and\ \bibinfo {author} {\bibfnamefont {H.}~\bibnamefont {Schmidt}},\ }\href
  {\doibase 10.1103/PhysRevB.81.165204} {\bibfield  {journal} {\bibinfo
  {journal} {Phys. Rev. B}\ }\textbf {\bibinfo {volume} {81}},\ \bibinfo
  {pages} {165204} (\bibinfo {year} {2010})}\BibitemShut {NoStop}%
\bibitem [{\citenamefont {Deng}\ \emph {et~al.}(2009)\citenamefont {Deng},
  \citenamefont {Tian}, \citenamefont {He}, \citenamefont {Bai}, \citenamefont
  {Xu}, \citenamefont {Yan}, \citenamefont {Dai}, \citenamefont {Chen},
  \citenamefont {Liu},\ and\ \citenamefont {Mei}}]{Deng2009}%
  \BibitemOpen
  \bibfield  {author} {\bibinfo {author} {\bibfnamefont {J.~X.}\ \bibnamefont
  {Deng}}, \bibinfo {author} {\bibfnamefont {Y.~F.}\ \bibnamefont {Tian}},
  \bibinfo {author} {\bibfnamefont {S.~M.}\ \bibnamefont {He}}, \bibinfo
  {author} {\bibfnamefont {H.~L.}\ \bibnamefont {Bai}}, \bibinfo {author}
  {\bibfnamefont {T.~S.}\ \bibnamefont {Xu}}, \bibinfo {author} {\bibfnamefont
  {S.~S.}\ \bibnamefont {Yan}}, \bibinfo {author} {\bibfnamefont {Y.~Y.}\
  \bibnamefont {Dai}}, \bibinfo {author} {\bibfnamefont {Y.~X.}\ \bibnamefont
  {Chen}}, \bibinfo {author} {\bibfnamefont {G.~L.}\ \bibnamefont {Liu}}, \
  and\ \bibinfo {author} {\bibfnamefont {L.~M.}\ \bibnamefont {Mei}},\ }\href
  {\doibase 10.1063/1.3206664} {\bibfield  {journal} {\bibinfo  {journal}
  {Appl. Phys. Lett.}\ }\textbf {\bibinfo {volume} {95}},\ \bibinfo {pages}
  {062513} (\bibinfo {year} {2009})}\BibitemShut {NoStop}%
\bibitem [{\citenamefont {Zhou}\ \emph {et~al.}(2009)\citenamefont {Zhou},
  \citenamefont {B\"urger}, \citenamefont {Helm},\ and\ \citenamefont
  {Schmidt}}]{Zhou2009}%
  \BibitemOpen
  \bibfield  {author} {\bibinfo {author} {\bibfnamefont {S.}~\bibnamefont
  {Zhou}}, \bibinfo {author} {\bibfnamefont {D.}~\bibnamefont {B\"urger}},
  \bibinfo {author} {\bibfnamefont {M.}~\bibnamefont {Helm}}, \ and\ \bibinfo
  {author} {\bibfnamefont {H.}~\bibnamefont {Schmidt}},\ }\href {\doibase 10.1063/1.3257363}
  {\bibfield  {journal} {\bibinfo  {journal} {Appl. Phys.
  Lett.}\ }\textbf {\bibinfo {volume} {95}},\ \bibinfo {pages} {172103}
  (\bibinfo {year} {2009})}\BibitemShut {NoStop}%
\bibitem [{\citenamefont {Morin}\ and\ \citenamefont {Maita}(1954)}]{Mor1954a}%
  \BibitemOpen
  \bibfield  {author} {\bibinfo {author} {\bibfnamefont {F.~J.}\ \bibnamefont
  {Morin}}\ and\ \bibinfo {author} {\bibfnamefont {J.~P.}\ \bibnamefont
  {Maita}},\ }\href {\doibase 10.1103/PhysRev.94.1525} {\bibfield  {journal}
  {\bibinfo  {journal} {Phys. Rev.}\ }\textbf {\bibinfo {volume} {94}},\
  \bibinfo {pages} {1525} (\bibinfo {year} {1954})}\BibitemShut {NoStop}%
\bibitem [{\citenamefont {Prince}(1953)}]{Pri1953}%
  \BibitemOpen
  \bibfield  {author} {\bibinfo {author} {\bibfnamefont {M.~B.}\ \bibnamefont
  {Prince}},\ }\href {\doibase 10.1103/PhysRev.92.681} {\bibfield  {journal}
  {\bibinfo  {journal} {Phys. Rev.}\ }\textbf {\bibinfo {volume} {92}},\
  \bibinfo {pages} {681} (\bibinfo {year} {1953})}\BibitemShut {NoStop}%
\bibitem [{\citenamefont {Riss}\ \emph {et~al.}(2009)\citenamefont {Riss},
  \citenamefont {Gerber}, \citenamefont {Korenblit}, \citenamefont {Suslov},
  \citenamefont {Passacantando},\ and\ \citenamefont {Ottaviano}}]{Riss2009}%
  \BibitemOpen
  \bibfield  {author} {\bibinfo {author} {\bibfnamefont {O.}~\bibnamefont
  {Riss}}, \bibinfo {author} {\bibfnamefont {A.}~\bibnamefont {Gerber}},
  \bibinfo {author} {\bibfnamefont {I.~Y.}\ \bibnamefont {Korenblit}}, \bibinfo
  {author} {\bibfnamefont {A.}~\bibnamefont {Suslov}}, \bibinfo {author}
  {\bibfnamefont {M.}~\bibnamefont {Passacantando}}, \ and\ \bibinfo {author}
  {\bibfnamefont {L.}~\bibnamefont {Ottaviano}},\ }\href {\doibase 10.1103/PhysRevB.79.241202}
  {\bibfield  {journal} {\bibinfo  {journal} {Phys.
  Rev. B}\ }\textbf {\bibinfo {volume} {79}},\ \bibinfo {pages} {241202}
  (\bibinfo {year} {2009})}\BibitemShut {NoStop}%
\bibitem [{\citenamefont {Fritzsche}\ and\ \citenamefont
  {Lark-Horovitz}(1959)}]{Fritzsche1959}%
  \BibitemOpen
  \bibfield  {author} {\bibinfo {author} {\bibfnamefont {H.}~\bibnamefont
  {Fritzsche}}\ and\ \bibinfo {author} {\bibfnamefont {K.}~\bibnamefont
  {Lark-Horovitz}},\ }\href {\doibase 10.1103/PhysRev.113.999} {\bibfield
  {journal} {\bibinfo  {journal} {Phys. Rev.}\ }\textbf {\bibinfo {volume}
  {113}},\ \bibinfo {pages} {999} (\bibinfo {year} {1959})}\BibitemShut
  {NoStop}%
\bibitem [{\citenamefont {Blakemore}(1962)}]{Bla1962}%
  \BibitemOpen
  \bibfield  {author} {\bibinfo {author} {\bibfnamefont {J.~S.}\ \bibnamefont
  {Blakemore}},\ }\href@noop {} {\emph {\bibinfo {title} {Semiconductor
  Statistics}}},\ edited by\ \bibinfo {editor} {\bibfnamefont {H.~K.}\
  \bibnamefont {Henisch}}\ (\bibinfo  {publisher} {Pergamon Press},\ \bibinfo
  {address} {New York},\ \bibinfo {year} {1962})\BibitemShut {NoStop}%
\bibitem [{\citenamefont {Woodbury}\ and\ \citenamefont
  {Tyler}(1955)}]{Woo1955}%
  \BibitemOpen
  \bibfield  {author} {\bibinfo {author} {\bibfnamefont {H.~H.}\ \bibnamefont
  {Woodbury}}\ and\ \bibinfo {author} {\bibfnamefont {W.~W.}\ \bibnamefont
  {Tyler}},\ }\href {\doibase 10.1103/PhysRev.100.659} {\bibfield  {journal}
  {\bibinfo  {journal} {Phys. Rev.}\ }\textbf {\bibinfo {volume} {100}},\
  \bibinfo {pages} {659} (\bibinfo {year} {1955})}\BibitemShut {NoStop}%
\bibitem [{\citenamefont {Hurd}(1972)}]{Hurd1972}%
  \BibitemOpen
  \bibfield  {author} {\bibinfo {author} {\bibfnamefont {C.~M.}\ \bibnamefont
  {Hurd}},\ }\href@noop {} {\emph {\bibinfo {title} {The Hall effect in metals
  and alloys}}}\ (\bibinfo  {publisher} {Plenum Press, New York},\ \bibinfo
  {year} {1972})\BibitemShut {NoStop}%
\bibitem [{\citenamefont {Solin}\ \emph {et~al.}(2000)\citenamefont {Solin},
  \citenamefont {Thio}, \citenamefont {Hines},\ and\ \citenamefont
  {Heremans}}]{Solin2000}%
  \BibitemOpen
  \bibfield  {author} {\bibinfo {author} {\bibfnamefont {S.~A.}\ \bibnamefont
  {Solin}}, \bibinfo {author} {\bibfnamefont {T.}~\bibnamefont {Thio}},
  \bibinfo {author} {\bibfnamefont {D.~R.}\ \bibnamefont {Hines}}, \ and\
  \bibinfo {author} {\bibfnamefont {J.~J.}\ \bibnamefont {Heremans}},\ }\href
  {\doibase 10.1126/science.289.5484.1530} {\bibfield  {journal} {\bibinfo
  {journal} {Science}\ }\textbf {\bibinfo {volume} {289}},\ \bibinfo {pages}
  {1530} (\bibinfo {year} {2000})}\BibitemShut {NoStop}%
\bibitem [{\citenamefont {Parish}\ and\ \citenamefont
  {Littlewood}(2005)}]{Parish2005}%
  \BibitemOpen
  \bibfield  {author} {\bibinfo {author} {\bibfnamefont {M.~M.}\ \bibnamefont
  {Parish}}\ and\ \bibinfo {author} {\bibfnamefont {P.~B.}\ \bibnamefont
  {Littlewood}},\ }\href {\doibase 10.1103/PhysRevB.72.094417} {\bibfield
  {journal} {\bibinfo  {journal} {Phys. Rev. B}\ }\textbf {\bibinfo {volume}
  {72}},\ \bibinfo {pages} {094417} (\bibinfo {year} {2005})}\BibitemShut
  {NoStop}%
\bibitem [{\citenamefont {Parish}\ and\ \citenamefont
  {Littlewood}(2003)}]{Parish2003}%
  \BibitemOpen
  \bibfield  {author} {\bibinfo {author} {\bibfnamefont {M.~M.}\ \bibnamefont
  {Parish}}\ and\ \bibinfo {author} {\bibfnamefont {P.~B.}\ \bibnamefont
  {Littlewood}},\ }\href {\doibase 10.1038/nature02073} {\bibfield  {journal}
  {\bibinfo  {journal} {Nature}\ }\textbf {\bibinfo {volume} {426}},\ \bibinfo
  {pages} {162} (\bibinfo {year} {2003})}\BibitemShut {NoStop}%
\bibitem [{\citenamefont {Hu}\ \emph {et~al.}(2007)\citenamefont {Hu},
  \citenamefont {Parish},\ and\ \citenamefont {Rosenbaum}}]{Hu2007}%
  \BibitemOpen
  \bibfield  {author} {\bibinfo {author} {\bibfnamefont {J.}~\bibnamefont
  {Hu}}, \bibinfo {author} {\bibfnamefont {M.~M.}\ \bibnamefont {Parish}}, \
  and\ \bibinfo {author} {\bibfnamefont {T.~F.}\ \bibnamefont {Rosenbaum}},\
  }\href {\doibase 10.1103/PhysRevB.75.214203} {\bibfield  {journal} {\bibinfo
  {journal} {Phys. Rev. B}\ }\textbf {\bibinfo {volume} {75}},\ \bibinfo
  {pages} {214203} (\bibinfo {year} {2007})}\BibitemShut {NoStop}%
\bibitem [{\citenamefont {Golikova}\ \emph {et~al.}(1962)\citenamefont
  {Golikova}, \citenamefont {Moizhes},\ and\ \citenamefont
  {Stil'bans}}]{Golikova1962}%
  \BibitemOpen
  \bibfield  {author} {\bibinfo {author} {\bibfnamefont {O.~A.}\ \bibnamefont
  {Golikova}}, \bibinfo {author} {\bibfnamefont {B.~Y.}\ \bibnamefont
  {Moizhes}}, \ and\ \bibinfo {author} {\bibfnamefont {L.~S.}\ \bibnamefont
  {Stil'bans}},\ }\href@noop {} {\bibfield  {journal} {\bibinfo  {journal}
  {Soviet Physics - Solid State}\ }\textbf {\bibinfo {volume} {3}},\ \bibinfo
  {pages} {2259} (\bibinfo {year} {1962})}\BibitemShut {NoStop}%
\bibitem [{Note1()}]{Note1}%
  \BibitemOpen
  \bibinfo {note} {In the case of a two-carrier-type model the same equations
  apply, when the sheet terms are replaced by their respective bulk
  counterparts.}\BibitemShut {Stop}%
\bibitem [{\citenamefont {Goldberg}\ and\ \citenamefont
  {Davis}(1956)}]{Goldberg1956}%
  \BibitemOpen
  \bibfield  {author} {\bibinfo {author} {\bibfnamefont {C.}~\bibnamefont
  {Goldberg}}\ and\ \bibinfo {author} {\bibfnamefont {R.~E.}\ \bibnamefont
  {Davis}},\ }\href {\doibase 10.1103/PhysRev.102.1254} {\bibfield  {journal}
  {\bibinfo  {journal} {Phys. Rev.}\ }\textbf {\bibinfo {volume} {102}},\
  \bibinfo {pages} {1254} (\bibinfo {year} {1956})}\BibitemShut {NoStop}%
\bibitem [{\citenamefont {Gallagher}\ and\ \citenamefont
  {Love}(1967)}]{Gallagher1967}%
  \BibitemOpen
  \bibfield  {author} {\bibinfo {author} {\bibfnamefont {J.~W.}\ \bibnamefont
  {Gallagher}}\ and\ \bibinfo {author} {\bibfnamefont {W.~F.}\ \bibnamefont
  {Love}},\ }\href {\doibase 10.1103/PhysRev.161.793} {\bibfield  {journal}
  {\bibinfo  {journal} {Phys. Rev.}\ }\textbf {\bibinfo {volume} {161}},\
  \bibinfo {pages} {793} (\bibinfo {year} {1967})}\BibitemShut {NoStop}%
\bibitem [{\citenamefont {Wick}(1954)}]{Wick1954}%
  \BibitemOpen
  \bibfield  {author} {\bibinfo {author} {\bibfnamefont {R.~F.}\ \bibnamefont
  {Wick}},\ }\href {\doibase 10.1063/1.1721725} {\bibfield  {journal} {\bibinfo
   {journal} {J. Appl. Phys.}\ }\textbf {\bibinfo {volume} {25}},\ \bibinfo
  {pages} {741} (\bibinfo {year} {1954})}\BibitemShut {NoStop}%
\bibitem [{\citenamefont {Jensen}\ and\ \citenamefont
  {Smith}(1972)}]{Jensen1972}%
  \BibitemOpen
  \bibfield  {author} {\bibinfo {author} {\bibfnamefont {H.~H.}\ \bibnamefont
  {Jensen}}\ and\ \bibinfo {author} {\bibfnamefont {H.}~\bibnamefont {Smith}},\
  }\href {\doibase 10.1088/0022-3719/5/20/006} {\bibfield  {journal} {\bibinfo
  {journal} {Journal of Physics C: Solid State Physics}\ }\textbf {\bibinfo
  {volume} {5}},\ \bibinfo {pages} {2867} (\bibinfo {year} {1972})}\BibitemShut
  {NoStop}%
\bibitem [{\citenamefont {Li}\ \emph {et~al.}(2006)\citenamefont {Li},
  \citenamefont {Wu}, \citenamefont {Guo}, \citenamefont {Luo},\ and\
  \citenamefont {Wang}}]{Li2006a}%
  \BibitemOpen
  \bibfield  {author} {\bibinfo {author} {\bibfnamefont {H.}~\bibnamefont
  {Li}}, \bibinfo {author} {\bibfnamefont {Y.}~\bibnamefont {Wu}}, \bibinfo
  {author} {\bibfnamefont {Z.}~\bibnamefont {Guo}}, \bibinfo {author}
  {\bibfnamefont {P.}~\bibnamefont {Luo}}, \ and\ \bibinfo {author}
  {\bibfnamefont {S.}~\bibnamefont {Wang}},\ }\href {\doibase 10.1063/1.2375015}
  {\bibfield  {journal} {\bibinfo  {journal} {J. Appl.
  Phys.}\ }\textbf {\bibinfo {volume} {100}},\ \bibinfo {eid} {103908}
  (\bibinfo {year} {2006})}\BibitemShut {NoStop}%
\bibitem [{\citenamefont {Zeng}\ \emph {et~al.}(2006)\citenamefont {Zeng},
  \citenamefont {Yao}, \citenamefont {Niu},\ and\ \citenamefont
  {Weitering}}]{Zen2006}%
  \BibitemOpen
  \bibfield  {author} {\bibinfo {author} {\bibfnamefont {C.}~\bibnamefont
  {Zeng}}, \bibinfo {author} {\bibfnamefont {Y.}~\bibnamefont {Yao}}, \bibinfo
  {author} {\bibfnamefont {Q.}~\bibnamefont {Niu}}, \ and\ \bibinfo {author}
  {\bibfnamefont {H.~H.}\ \bibnamefont {Weitering}},\ }\href {\doibase 10.1103/PhysRevLett.96.037204}
  {\bibfield  {journal} {\bibinfo  {journal}
  {Phys. Rev. Lett.}\ }\textbf {\bibinfo {volume} {96}},\ \bibinfo {pages}
  {037204} (\bibinfo {year} {2006})}\BibitemShut {NoStop}%
\bibitem [{\citenamefont {Zeng}\ \emph {et~al.}(2008)\citenamefont {Zeng},
  \citenamefont {Zhang}, \citenamefont {van Benthem}, \citenamefont
  {Chisholm},\ and\ \citenamefont {Weitering}}]{Zeng2008}%
  \BibitemOpen
  \bibfield  {author} {\bibinfo {author} {\bibfnamefont {C.}~\bibnamefont
  {Zeng}}, \bibinfo {author} {\bibfnamefont {Z.}~\bibnamefont {Zhang}},
  \bibinfo {author} {\bibfnamefont {K.}~\bibnamefont {van Benthem}}, \bibinfo
  {author} {\bibfnamefont {M.~F.}\ \bibnamefont {Chisholm}}, \ and\ \bibinfo
  {author} {\bibfnamefont {H.~H.}\ \bibnamefont {Weitering}},\ }\href {\doibase 10.1103/PhysRevLett.100.066101}
  {\bibfield  {journal} {\bibinfo  {journal}
  {Phys. Rev. Lett.}\ }\textbf {\bibinfo {volume} {100}},\ \bibinfo {pages}
  {066101} (\bibinfo {year} {2008})}\BibitemShut {NoStop}%
\bibitem [{\citenamefont {Tsuchida}\ \emph {et~al.}(2009)\citenamefont
  {Tsuchida}, \citenamefont {Asubar}, \citenamefont {Jinbo},\ and\
  \citenamefont {Uchitomi}}]{Tsuchida2009}%
  \BibitemOpen
  \bibfield  {author} {\bibinfo {author} {\bibfnamefont {R.}~\bibnamefont
  {Tsuchida}}, \bibinfo {author} {\bibfnamefont {J.~T.}\ \bibnamefont
  {Asubar}}, \bibinfo {author} {\bibfnamefont {Y.}~\bibnamefont {Jinbo}}, \
  and\ \bibinfo {author} {\bibfnamefont {N.}~\bibnamefont {Uchitomi}},\ }\href
  {\doibase 10.1016/j.jcrysgro.2008.09.112} {\bibfield  {journal} {\bibinfo
  {journal} {Journal of Crystal Growth}\ }\textbf {\bibinfo {volume} {311}},\
  \bibinfo {pages} {937} (\bibinfo {year} {2009})}\BibitemShut {NoStop}%
\bibitem [{\citenamefont {Park}\ \emph {et~al.}(2001)\citenamefont {Park},
  \citenamefont {Wilson}, \citenamefont {Hanbicki}, \citenamefont {Mattson},
  \citenamefont {Ambrose}, \citenamefont {Spanos},\ and\ \citenamefont
  {Jonker}}]{Par2001}%
  \BibitemOpen
  \bibfield  {author} {\bibinfo {author} {\bibfnamefont {Y.~D.}\ \bibnamefont
  {Park}}, \bibinfo {author} {\bibfnamefont {A.}~\bibnamefont {Wilson}},
  \bibinfo {author} {\bibfnamefont {A.~T.}\ \bibnamefont {Hanbicki}}, \bibinfo
  {author} {\bibfnamefont {J.~E.}\ \bibnamefont {Mattson}}, \bibinfo {author}
  {\bibfnamefont {T.}~\bibnamefont {Ambrose}}, \bibinfo {author} {\bibfnamefont
  {G.}~\bibnamefont {Spanos}}, \ and\ \bibinfo {author} {\bibfnamefont {B.~T.}\
  \bibnamefont {Jonker}},\ }\href {\doibase 10.1063/1.1369151} {\bibfield
  {journal} {\bibinfo  {journal} {Appl. Phys. Lett.}\ }\textbf {\bibinfo
  {volume} {78}},\ \bibinfo {pages} {2739} (\bibinfo {year}
  {2001})}\BibitemShut {NoStop}%
\bibitem [{\citenamefont {Yu}\ \emph {et~al.}(2010)\citenamefont {Yu},
  \citenamefont {Jamet}, \citenamefont {Devillers}, \citenamefont {Barski},
  \citenamefont {Bayle-Guillemaud}, \citenamefont {Beign{\'e}}, \citenamefont
  {Rothman}, \citenamefont {Baltz},\ and\ \citenamefont {Cibert}}]{Yu2010}%
  \BibitemOpen
  \bibfield  {author} {\bibinfo {author} {\bibfnamefont {I.-S.}\ \bibnamefont
  {Yu}}, \bibinfo {author} {\bibfnamefont {M.}~\bibnamefont {Jamet}}, \bibinfo
  {author} {\bibfnamefont {T.}~\bibnamefont {Devillers}}, \bibinfo {author}
  {\bibfnamefont {A.}~\bibnamefont {Barski}}, \bibinfo {author} {\bibfnamefont
  {P.}~\bibnamefont {Bayle-Guillemaud}}, \bibinfo {author} {\bibfnamefont
  {C.}~\bibnamefont {Beign{\'e}}}, \bibinfo {author} {\bibfnamefont
  {J.}~\bibnamefont {Rothman}}, \bibinfo {author} {\bibfnamefont
  {V.}~\bibnamefont {Baltz}}, \ and\ \bibinfo {author} {\bibfnamefont
  {J.}~\bibnamefont {Cibert}},\ }\href {\doibase 10.1103/PhysRevB.82.035308}
  {\bibfield  {journal} {\bibinfo  {journal} {Phys. Rev. B}\ }\textbf {\bibinfo
  {volume} {82}},\ \bibinfo {pages} {035308} (\bibinfo {year}
  {2010})}\BibitemShut {NoStop}%
\end{thebibliography}
\end{document}